# NSUN6 Promotes breast cancer Progression via m5C-Dependent Regulation of Proliferation and Migration Networks

**Authors** Zhengyu Song
**Affiliation** Department of Biomedical Engineering, The Hong Kong Polytechnic University, Hong Kong, 999077
**Email**:zhengyu.song@connect.polyu.hk

**Abstract**

**Background:** RNA 5-methylcytosine (m5C) modification is emerging as a critical regulator of cancer progression. NSUN6 is an m5C methyltransferase with unclear functional significance in breast cancer.

**Purpose:** This study aimed to elucidate the oncogenic role of NSUN6 and its downstream molecular network in breast cancer cells,with a focus on proliferation and migration—key drivers of tumor progression.

**Methods:** Stable NSUN6 knockdown and overexpression models were established in MCF7 cells using lentiviral transduction. Phenotypic effects were assessed     by CCK-8 assay, real-time imaging, and Transwell assay. Transcriptomic profiling (RNA-seq) and bioinformatic analyses were employed to identify downstream targets and pathways.

**Results:** NSUN6 knockdown significantly suppressed cell proliferation, while its overexpression enhanced of proliferation and migration. RNA-seq revealed hundreds of differentially expressed genes influenced by NSUN6.   Ten candidate genes, including EFEMP1, TNS3, and FGFR2 showed expression levels positively correlated with NSUN6, forming a potential regulatory network linked to cancer-relevant processes.

**Conclusion:** NSUN6 drives breast cancer cell proliferation and migration, likely through m5C-mediated regulation of a distinct genes set. These results position NSUN6 as a novel epigenetic regulator and potential therapeutic target in breast cancer, providing a mechanistic basis for further translational investigation.

# Introduction

RNA modification represents a series of finely regulated molecular processes **that influence** RNA metabolism, including splicing, translation, localization, stability, turnover, and interactions with RNA-binding proteins (RBPs) or other RNAs, thereby diversifying genetic information [1]. Over 170 different types of RNA modifications have been identified, and these modifications are widely present in all types of RNA, greatly increasing the functionality of RNA molecules [2]. Similar to epigenetics, a group of proteins have been identified that specifically 'writer' (catalyze the deposition of a specific modification), 'eraser' (catalyze the removal of a specific modification), and 'reader' (recognize and bind modified nucleotides) to influence RNA fat [3]. At present, research on these modification functions is emerging. Recent developments in high-throughput sequencing and improved techniques such as mass spectrometry have shed light on the exciting new field of RNA surface transcriptomics. Researchers have revealed many pathways by which RNA modification affects RNA metabolism and have been shown to have a large impact on human pathology.

6-methyladenosine (m6A) is the most abundant and well-characterized internal modification in mRNA. Of all transcripts encoded by mammalian cells, 20-40% are m6A methylated, and methylated mRNA tends to contain multiple m6A per transcript [2,4]. Its role is to regulate the self-renewal of embryonic stem cells and cancer cells. This is because the m6A marker encodes transcripts of important developmental regulators to facilitate their turnover during the transition of cell fate, thus enabling the cell to properly switch its transcriptome from one cell state to another [5].

M5C modification is a widely distributed RNA modification in eukaryotes. Studies in recent years have found that m5C modification exists on tRNA, rRNA, and mRNA [6,7]. Currently, researchers have found that there are two types of m5C sites in mRNA (named Type I and Type II). Type I sites are enriched in most tissues and cell types according to their unique sequence and structural characteristics [8,9]. Type I sites are located at the 5th end of a stem-loop structure, and there is a g-rich motif at the 3rd end of the site. The writer protein of site I is NSUN2, a writer protein that can also mediate the methylation of m5C at the end of the tRNA T $\psi$ C ring [8]. Type II site has strong tissue specificity, accounting for 1%-40%, and Type II site is located in the ring region of a stem-ring structure, and site 3 terminating a TCCA motif Type II site was NSUN6 [10]. In addition, m5C is a conserved and universal marker of RNA in all life domains. Current detection methods for m5C rely on the chemical reactivity of cell pyrimidines in the presence of sodium bisulfite, or immunoprecipitation using antibodies against m5C, or RNA methyltransferases previously cross-linked to RNA targets. RNA bisulfite sequencing is the most used technique for mapping m5C, and it also effectively detects m5C in rich RNA such as tRNA and rRNA [2,4].

Previous studies have identified the NSUN6 as an m5C methyltransferase targeting mRNA, which is possible as part of the quality control mechanism involved in the fidelity of translation termination [8,11]. And the other group also found that ribosomal analysis of NSUN6-targeted regulation of mRNA methylation further demonstrated that NSUN6-specific methylation was associated with translation termination. Although NSUN6 is not essential for embryonic development in mice, it is downregulated in human tumors, and some cancer patients with high NSUN6 expression have a better prognosis [11]. In addition, how to find and sequence NSUN6-mediated m5C modification sites is also an important topic. At present, relatively mature techniques include CRISPR integrated gRNA and reporter sequencing (CIGAR-Seq), and methylation-dependent individual-nucleotide resolution cross-linking and immunoprecipitation (miCLIP) binding RNA bisulfite sequencing. These technologies effectively helped us to search for RNA-dependent NSUN6 m5C sites in the human transcriptome [8,11].

According to a study of global cancer statistics, breast cancer is the most commonly diagnosed cancer in women, and one of the leading causes of cancer-related death younger women each year are at risk, greater risk in developing countries [12]. Therefore, despite the advanced treatment and early detection improves the survival rate, the pathogenic factors of breast cancer are still not clear. Based on the expression of hormone receptors (estrogen (ER) progesterone (PR)) and human epidermal growth factor receptor 2 (HER2), breast cancer has four molecular subtypes, namely luminal A, luminal B, HER2-positive and basal-like or triple-negative (TNBC) [12,13]. The HER2 subtype is ER and PR negative but positive for HER2 and the TNBC presents a triple-negative immunophenotype (ER, PR, and HER2 negative), increased proliferation rate, and the highest incidence of relapse [14,15]. Recently, Huang et al. reported that all eleven m5C regulators (NSUN2-NSUN7, DNMT1, DNMT3A, DNMT3B, ALYREF, and TET2) were differentially expressed in TNBC and can potentially predict clinical prognostic risk in patients. The up-regulation of NSUN6 expression is closely related to cell cycle signaling, RNA degradation, and RNA polymerase, while the down-regulation of NSUN6 expression is related to cell adhesion, metabolism, and extracellular matrix receptor interactions [12]. Michigan Cancer Foundation-7 (MCF7) cell line is a non-triple-negative breast cancer model, which is often used in the study of breast cancer transplanted tumors. It has been found in previous studies that m5C affects tumor development in patients and the influence of tumor immune microenvironment and potential markers. For breast cancer, identifying the function of RNA modification in its development may help us find relevant pathogenesis and treatment methods [14,16]. However, the precise functional impact of NSUN6 in breast cancer and its underlying molecular mechanisms remain poorly characterized, limiting our understanding of its potential as a therapeutic target. Specifically, whether and how NSUN6 regulates the aggressive phenotypes of breast cancer cells through its m5C methyltransferase activity is unknown. Here, we combined genetic perturbation, phenotypic characterization, and transcriptomic profiling to systematically investigate the oncogenic role of NSUN6 in breast cancer. We demonstrate that NSUN6 drives proliferation and migration in MCF7 cells and identify a network of downstream genes—including EFEMP1, TNS3, and FGFR2—that are positively correlated with

NSUN6 expression and linked to key cancer pathways. This study not only elucidates a novel epigenetic mechanism promoting breast cancer progression but also nominates NSUN6 as a potential druggable target, providing a rationale for future development of epitranscriptome-directed therapies in oncology.

# MATERIALS AND METHODS

## Cell culture

Michigan Cancer Foundation-7 (MCF7), HEK293 (HEK), and MDA-MB-231 (M231) were obtained from the ATCC and were grown in DMEM media (Thermo Fisher Scientific) supplemented with 1 mM glutamax (Thermo Fisher Scientific), 10% heat-inactivated FBS (Thermo Fisher Scientific). Cells were cultured in an incubator (Thermo Fisher Scientific) with 5% $CO_2$ and constant 37°C. During cell passage, 0.25% Trypsin-EDTA (1×, Thermo Fisher Scientific) was used to digest the cells and 1× PBS (mich SCIENTIFIC) to wash impurities.

## Knockdown and overexpression of NSUN6 in MCF7

The design of four NSUN6-shRNA oligos (sequences:
oligo-hNSUN6-shRNA1-Top
CCGGCGTCTGGCTAATAAGGACTCTCTCGAGAGAGTCCTTATTAGCCAGACGTTTTTG;
oligo-hNSUN6-shRNA1-Bottom
AATTCAAAAACGTCTGGCTAATAAGGACTCTCTCGAGAGAGTCCTTATTAGCCAGACG;
oligo-hNSUN6-shRNA2-Top
CCGGGCAAAGAAATCTTCAGTGGATCTCGAGATCCACTGAAGATTTCTTTGCTTTTTG;
oligo-hNSUN6-shRNA2-Bottom
AATTCAAAAAGCAAAGAAATCTTCAGTGGATCTCGAGATCCACTGAAGATTTCTTTGC) and NSUN6 overexpression vectors. The sequence of the target gene was obtained from NCBI (http://www.ncbi.nlm.nih.gov/ Gene ID: 221078) and vectors were designed online (http://crispr.mit.edu/, https://design.synthego.com, snapgene) and checked the specificity of the shRNA online (NCBI BLAST). The PLKO vectors were digested by AGEI and EcoRI (NEB) and then purified by kit. (VAZYME-RM201. gel extraction kit). The shRNA oligos were annealed by protocol and cloned the shRNA into plko vectors to recombine. (VAZYME-C112. ClonExpress II One Step Cloning Kit). Then it was transformed the recombinant plasmids into DH5α Chemically Competent Cell (KT HEALTH, KTSM101L) and cultured them in LB/Amp solid medium. 16 hours later, we selected positive colonies and sequenced them. We constructed the overexpression vector lentivirus-NSUN6-T2A-puro, which amplified NSUN6, T2A, and GFP fragments and assembled clones. The PCR product was purified by gel electrophoresis (Regular Agarose G-10, Biowest; TAE, 50x, Beyotime, ST716, diluted into 0.5x for use) and gel extraction (FastPure Gel DNA Extraction Mini Kit, Vazyme, DC301). Similarly, the recombinant plasmids were cultured in LB/Amp solid medium. After 16 hours, positive colonies were selected and sequenced. Isolated the plasmid DNA with a plasmid DNA mini kit (TIANGEN -DP118. EndoFree Mini Plasmid Kit II). Plasmid DNA isolation without endotoxin and were determined the sample concentration by UV-VIS spectrophotometer (NanoDrop 2000). The PCR Primers were designed online: (https://design.synthego.com, snapgene) and PCR reagents (Vazyme Phanta Max Super-fidelity DNA polymerase Kit) sequence analysis (snapgene). Finally, we obtained two shNSUN6 plasmids and their control, and two overexpressed plasmids and their control.

MCF7 cell lines with stable knockdown of NSUN6 were cultured using lentivirus transfection. The lentivirus was packaged in HEK293T with the helper plasmids psPAX2 (Addgene, #12259) and pMD2.G (Addgene, #12259) to construct the stable overexpression cell line. And the reagents used polyethyleneimine (PEI) as a transfection medium. Before transfection, we inoculated HEK-293T cells in a 6-well plate (105 cells/well, 2mL DMEM/well) for 24 hours. Then, we premixed 1ug target plasmid, 1.5ug psPAX2 and 1ug pMD2.G with 200µL opti-MEM (Thermo Fisher Scientific) and 10.5ul PEI (1 µg/L, Polysciences) together and incubated for 15 minutes. Then, we transferred the mix to the HEK-293T cells. Cultured at 37°C with 5% carbon dioxide for 48h, we collected the supernatant and centrifuged it for 10 minutes at 2000 g to remove cell debris and impurities. Then, the product was filtrated through a 0.45µm filter (Millex, #SLHV033RB). The final lentivirus was separated into 1.5mL centrifuge tubes for 1mL each. The virus particles were stocked in -80°C freezer for further use.

For transduction, we added 1mL specific virus particles into MCF7 and cultured the cells under the same conditions for 3-5 days to obtain stable MCF7 cell lines with low NSUN6 knockdown. After incubation at 37°C for 24 hours, we replaced the DMEM medium and added the antibiotics puromycin (1 µg/ml) to select targeting cells. The overexpression vector that transfected to MCF7 cell line also used lentivirus transfection.

## Cell sorting

The cells were sorted by flow cytometry. We divide the cells into fluorescent and non-fluorescent parts and determine the proportion of different divisions to determine the rate of transfection. The cell without fluorescing is singled out and will be checked under the microscope. The single cell then will be cultured for further experiments.

## Migration assay

NSUN6 overexpression MCF7 cells and normal control were collected for Migration assay. First, the cell suspension was prepared, and after washing with PBS, 0.25% trypsin- EDTA(Gibco-25200072) was digested, and digestion was terminated. The cell count was carried out. According to the cell density, the quantity of 12/180,000 cells was taken and diluted to 300/450ul. During this assay, cells were cultured on the upper layer of an insert with a permeable membrane in a well of a 12-well plate. Transwell Insert (Costar, 8um) was used to inoculate the cells, and a 600ul DMEM culture medium containing 20%FBS was added into the middle part of the lower chamber. A circle of PBS was placed around the side to prevent water evaporation. 150ul cell suspension was added to the upper chamber and cultured at 37℃ with 5% carbon dioxide for 48h. When collected the transwell plate, we fixed it with methanol, dyed it with 0.5% crystal violet dye for 30 minutes, rubbed the tip of the cotton swab to make it shaggier, rotated to wipe off the excess cells on the membrane, opened the cover and dried it, quantified it with the microscope (Nikon, #Ts2-FL) the next day, and analyzed and counted it with ImageJ software. Each transwell filter takes three stable images.

## Cell proliferation

We collected logarithmic growth phase MCF7 cells with normal growth states and inoculated them into 96-well plates at the rate of 4000 cells per well. 8 multiple Wells were set in each group and placed in a cell incubator for culture. After cells were attached to the wall, they were transfected with the shRNA expression vector and NSUN6 overexpression vectors and then cultured for 48h Then, discard the drug-containing medium, add the freshly prepared toxicity detection solution CCK8 containing 10uL into each well, and place it in the incubator for further cultivation for 4h. Then, use the microplate to measure the OD value of 450 nm and repeat the experiment 3 times. The mean value of the experimental results was taken as the experimental result and the growth inhibition rate was calculated by the formula = [(control group OD - experimental group OD)/ control group OD] 100%. The histogram of cell growth inhibition rate was drawn with hour as abscissa and fold change of growth difference as ordinate.

## RNA extraction

For the stable production of MCF7 cell lines after transfection, we harvested NSUN6 knockdown MCF7 cell lines that were already covered with six-hole plates and extracted total RNA from them using column extraction (VAZYME-RC101.FastPure Cell/Tissue Total RNA Isolation Kit). Then we perform gel electrophoresis to test the quality of RNA, and the loading RNA sample has been denatured at 65°C for five minutes. Finally, we obtained a relatively complete MCF7 cell total RNA.

## RT-qPCR

For the stable production of MCF7 cell lines after transfection, we isolated the RNA by column extraction (VAZYME-RC101.FastPure Cell/Tissue Total RNA Isolation Kit). Then we perform gel electrophoresis to test the quality of RNA. Then we performed reverse transcription to get the DNA of the cells by the kit (YESEN-11121ES60.Hifair® II 1st Strand cDNA Synthesis Kit (gDNA digester plus)). For each sample, we used 1μg extracted RNA to do the transcription. Then, we diluted the cDNA samples into 1:20 to do the RT-qPCR. The cDNA will be used for quantitative reactions. The PCR Primers were designed online: (https://design.synthego.com, snapgene). We used the GAPDH gene as the reference gene and select the qPCR TaqMan probes (YESEN-11211ES03.Hieff Unicon® Universal TaqMan multiplex qPCR master mix). And we performed qPCR to get the expression level of the NSUN6 gene in a different group of cells by the delta CT method.

## DNA extraction

We harvested NSUN6-knockdown MCF7 cell lines that were already covered with six-hole plates, and we extracted genomic DNA from them using a column. (VAZYME-DC112.FastPure Blood/Cell/Tissue /Bacteria DNA Isolation Mini Kit) Then we performed PCR and gel electrophoresis for the target NSUN6 fragment. And then we proved the target DNA bands and performed sequence analysis.

## Western blot

We used the RIPA buffer to extract total proteins. Cells were rinsed with outdoor PBS, lysed in cold RIPA buffer (50 mM Tris HCl pH 7.4, 1% NP-40, 150 mM NaCl, 0.1% SDS, 1% Triton x-100,1% sodium deoxycholate). RIPA was supplemented with a protease inhibitor mixture (100 x) PIC (11836170001, Roche). Cells were collected with a cell scraper, and the mixture was centrifuged for 15 min at maximum speed in a pre-cooled 4°C centrifuge. The supernatant was taken for the determination of protein concentration and denaturation of WB samples. Cell protein lysate and upper sample buffer (5X) (Beyotime-P0015L) were added into each pore of 10% SDS polyacrylamide gradient gel after denaturation at 95°C for five minutes, and a protein marker was added. For upper stacking gel, the samples were run by gel electrophoresis for 30 minutes until all the samples were on the same level. For the separation gel, the samples were run by electrophoresis for 60 minutes. After electrophoresis, the protein was transferred to the polyvinylidene difluoride (PVDF) membrane (Healthcare) under gel electrophoresis for 100 minutes and the PVDF membrane was blocked with 10% non-fat milk (BBI, #A600669) at room temperature. Then the blocked PVDF membranes were washed with TBST buffer 3 times, 10 minutes each. 1X TBST buffer was diluted from 10x TBS buffer (20mM pH7.6 Tris-HCl buffer, 150mM NaCl for 1L ddH2O) and 0.1% Tween 20) containing 5% (W/V) skim milk. The primary antibody was diluted in the same milk containing 1X TBST solution and we incubated the membranes with specific primary antibodies for 12 hours at 4°C after cutting the membranes into specific pieces. Then the membrane was washed three times with the previous TBS-T for 10 minutes each time, and Appropriate horseradish peroxidase (HRP) labeled secondary antibody (1:10 000) was incubated in TBS-T at room temperature for 1 h. After the last washing, we used the western HRP substrate detection reagent (Immobilon Western HRP-WBKLS 0100) to fix and develop the image. Proteins were then detected using the PierceIII ECL Western Blotting Substrate (Thermo, #32209) on BIO-RAD ChemiDocTM XRS+ system. The primary antibody is NSUN6 (ABclonal-A7205. NSUN 6 Rabbit pAb), GAPDH (ProteinTech-Cat No. 60004-1-IG). Resistance to Anti-Rabbit (ABclonal-AS014), Anti-Mouse (ABclonal-AS003).

## mRNA library construction and RNA-seq analysis

We isolated the total RNA of MCF7 cells and separated the mRNA fragments, and then synthesized the first and second cDNA strands respectively. For each sample, 5ug total RNA was used as starting material. After the adapt attachment and enrichment of the library, we performed high-throughput sequencing of the product. The method and reagent are from VAHTS reactive mRNA- Seq Library Prep Kit for Illumina V2 (Vazyme, NR612). Raw pair-end reads were trimmed by using cutadapt (-a AGATCGG AAGAG -A AGATCGGAAGAG -j 8). Cleaned reads were mapped to the human reference genomes (GRCH37) by hisat2 (default parameters). All mapping results were counted by featureCounts (-t exon -s 2 -B -C, annotation file is GRCm38.102). DEGs were calculated by DEseq2 (fc>2, fdr<0.1).

## GO enrichment analysis

GO enrichment was performed by the cluster profile of R package. For each tissue type, expressed genes were used as the background.

## Results

## Knockdown of NSUN6 could suppress MCF7 cell proliferation in MCF7

Some previous studies have found that Nsun6 can inhibit the proliferation of HCC cells, and cell proliferation and migration are usually mutually controlled [17,18]. We successfully transfected four plasmids shCon-1, shCon-2, shNSUN6-1, and shNSUN6-2 into MCF7 cells by lentivirus transfection. We obtained successfully transfected cells through drug screening and observed whether the expression of NSUN6 protein decreased by extracting total cell protein and conducting Western blotting. Compared with the control group, the expression level of NSUN6 protein was significantly decreased in both shNSUN6-1 and shNSUN6-2 groups (Figure 1B), indicating that we successfully constructed stable MCF7 cell lines with NSUN6-inhibited expression. At the same time, we also carried out cell proliferation assay and measured the proliferation of MCF7 cells in each group at different time periods. The fluorescence intensity data of time-lapse imaging were normalized by dividing the fluorescence intensity at each time point (I) by the fluorescence intensity at the first time point ($I_0$). The image shows that when the expression of NSUN6 is inhibited, the cell proliferation of MCF7 will also be significantly decreased after 40 hours (Figure 1A). Therefore, we conclude that NSUN6 knockdown can inhibit cell proliferation in MCF7.

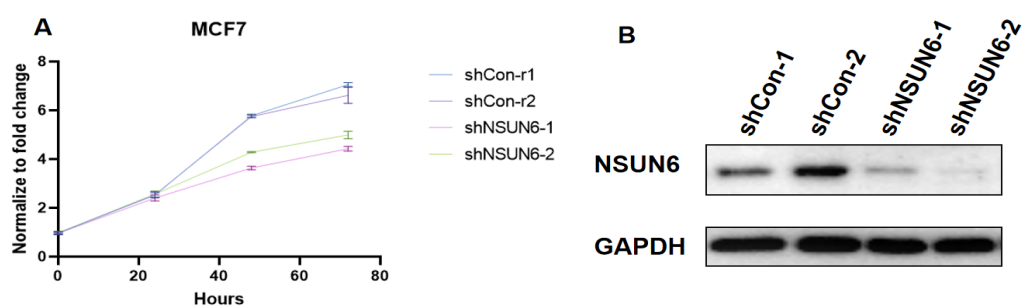

**Figure 1.** Inhibition of NSUN6 expression could affect MCF7 cell proliferation. (A) MCF7 cell proliferation in each group was measured at different time periods and normalized cell proliferation diagram. (B) Western blot has obtained the results showed that NSUN6 protein expression was knockdown in two MCF7-shnusn6 cell lines.

## NSUN6 overexpression could enhance cell proliferation and migration in MCF7.

In order to study the effects of excessive NSUN6 on the proliferation and migration of MCF7 cell lines, we constructed two types of overexpression vectors with control groups, one carrying puromycin resistance (OE-puro, OE-NSUN6-T2A-puro) and one carrying green fluorescent protein resistance (OE-GFP, OE-NSUN6-linker-GFP). Also performed by the lentivirus transfection method, these plasmids were transferred into normal MCF7 cell lines. MCF7 cell lines with stable overexpression of NSUN6 were obtained by drug screening and flow cytometry analysis. We performed cell proliferation under the same conditions as before. After normalization to fold change, it was found that the growth trend of cells overexpressing NSUN6 protein in both groups increased significantly compared with the control group after 40 hours (Figure 2A–B). We observed a more significant difference in cell proliferation in the GFP group. Then, the total protein of MCF7 cells was extracted for Western blot to observe whether the expression of NSUN6 protein increased. Compared with the control group, NSUN6 overexpressed cellular protein showed two bands, and the one with higher molecular weight was identified as the overexpressed NSUN6-GFP protein, indicating that the expression level of NSUN6 was indeed higher than that of the control group (Figure 2C). Therefore, we selected a part of MCF7 cell lines with GFP resistance whose cell proliferation changed greatly for the cell migration experiment. After migration for 72 h, we found that the migration ability of MCF7 cell overexpressing NSUN6 was significantly stronger than that of the control group after fixation and staining (Figure 2D–E). These results demonstrated that NSUN6 overexpression could enhance the proliferation and migration of MCF7 cells, and indirectly indicate that there is a certain connection between cell growth and migration.

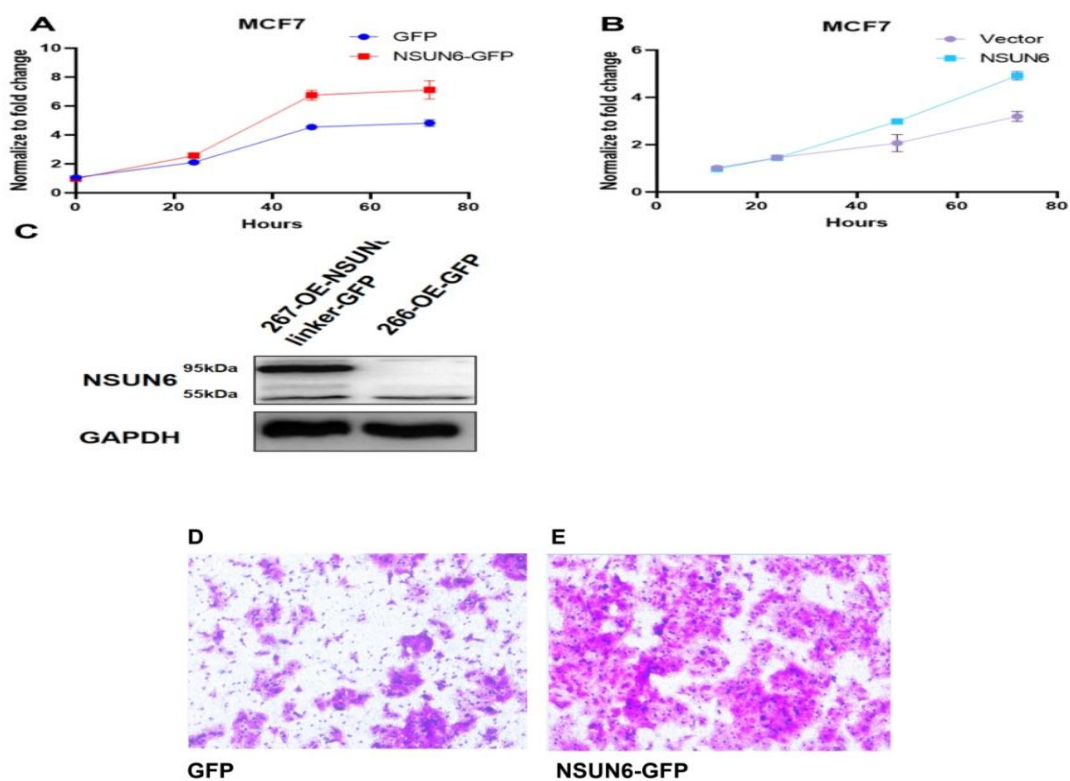

**Figure 2.** NSUN6 overexpression enhanced the proliferation and migration of MCF7 cells. (A, B) The proliferation of MCF7 cells at different time periods in each group was measured and normalized cell proliferation maps were drawn. (A) Overexpression vectors labeled with green fluorescent protein and control were introduced. (B) Overexpression vectors labeled with puromycin and control were introduced. (C) Western blot results showed that NSUN6 protein was overexpressed in MCF7 cell lines, and the molecular size of the overexpression band was larger than the normal NSUN6 band. (D, E) Results of migration experiment on MCF7 cells with GFP marker group and imaged by the microscope after 72h.

## RNA-seq data revealed DEGs and potential regulated genes related with cell proliferation

To explore the mechanism of how NSUN6 affected cell proliferation in MCF7, we extracted total RNA from MCF7 cell lines with NSUN6 overexpression and knockdown, and then we performed RNA-seq. We found that 381 genes were up-regulated, and 152 genes were down-regulated in the transcriptome of MCF7 cell lines with knockdown NSUN6 expression (Figure 3C). And we performed GO enrichment analysis on the genes that differ in these changes, explored the biological pathways in which these genetic changes are concentrated (Figure 3A, 3B). From the results, we observed in the up-regulated bubble diagram that the genes were concentrated in biological pathways such as innate immune response, MAPK cascade activation regulation, and cytokine production, and the P-value was less than 0.00075. In the bubble map of downregulated, these genes were concentrated in the biological pathways such as embryonic morphology, brain development, and developmental growth, but their P values were all greater than 0.006. At the same time, compared with the control group, we found that in the transcriptome of the MCF7 cell line overexpressing NSUN6, 53 genes were up-regulated, and 50 genes were down-regulated (Figure 3F). We also performed GO analysis on NSUN6 overexpressed RNA-seq data. Through the bubble diagram, it was found that the gene up-regulation caused by NSUN6 overexpression was concentrated in the extracellular matrix and extracellular matrix containing collagen, with the P-value greater than 0.01. The downregulation of NSUN6 overexpression was concentrated in the process of heart development and muscle structure development, with the P-value less than 0.048 (Figure 3E, 3D). And then,by comprehensively comparing the DEG data of MCF7_sh and MCF7_OE, we found that a total of 539 genes were up-regulated or down-regulated. The map selected the absolute value of log$_2$FC between the two genes closing to 1 as the differential genes, and there were about 43 common differential genes (Figure 3G). NSUN6 in the lower-left corner of the figure indicates that it is knocked down by 2.4 times on the abscissa and overexpressed by 4 times on the ordinate. As previous phenotypic experiments have verified that knockdown of NSUN6 inhibits cell proliferation, while overexpression of NSUN6 promotes cell proliferation, we selected genes in a half region of NSUN6 as potential target genes related to cell proliferation (Figure 3G). There are about 10 genes in this part, namely EFEMP1, TNS3, FGFR2, PGM5, SCIN, LINC01671, HAPLN1, SYNM, DEGS1, and ATP7B (Figure 3G). In some previous studies, EFEMP1, TNS3, and FGFR2 were associated with cell proliferation [19–21]. In short, these results showed that the change of NSUN6 expression could up-regulate or down-regulate some modified genes, and thereby affect the normal growth of cells, which indicated that these differentially expressed genes may influence cell proliferation or migration.

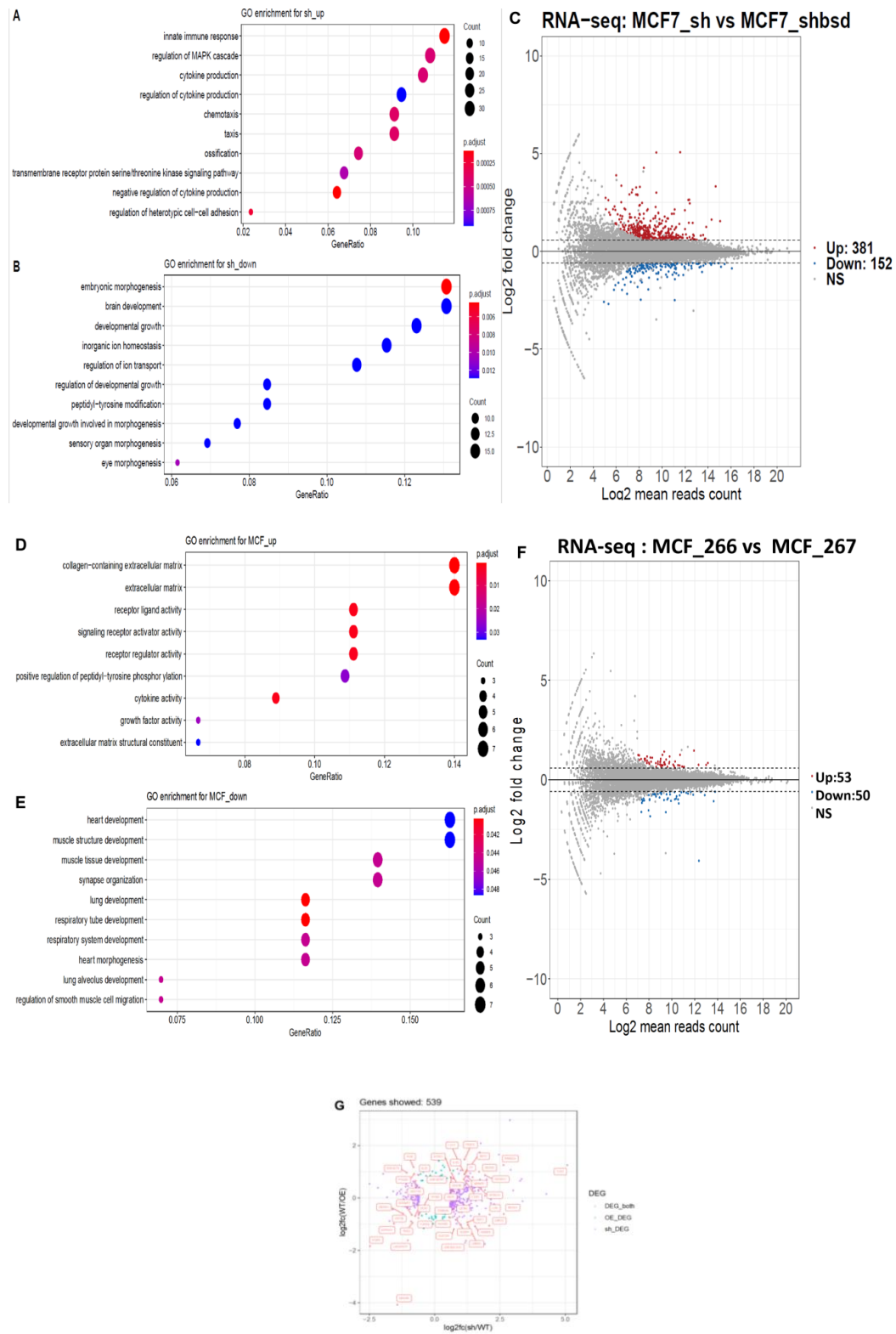

**Figure 3.** RNA-seq was used to analyze the effects of NSUN6 knockdown and overexpression on m5C-related modification genes. (A, B) Enrichment GO analysis

was performed on the knockdown of NSUN6, and the analysis results of the biological process were displayed by the bubble diagram, and the ratio of the number of genes related to this Term in each differential gene to the total number of differential genes was obtained. (C) Images of changes in expression levels of related genes after down-regulation of NSUN6 based on RNA-seq difference analysis, |FC| > 1.5 FDR = 0.05. (D, E) Enrichment GO analysis was performed for the overexpressed NSUN6 gene, and the ratio of the number of genes related to this Term in each differential gene to the total number of differential genes was obtained by bubble map showing the results of biological process analysis. (F) Images of changes in related gene expression levels after NSUN6 upregulation based on RNA-seq difference analysis, |FC| > 1.5, FDR = 0.05. This data is about transfection of plasmids 266 and 267, which 266 plasmid was only labeled with a green fluorescent protein, and 267 plasmids were overexpressed NSUN6 plasmids. (G) By comparing the DEG data of MCF7_sh and MCF7_OE, a total of 539 genes showed changes in expression, and the absolute value of log2FC near 1 between the two data was selected as the comment differential gene.

# Discussions

## Clinical significance of m5C modification and NSUN6 in breast cancer

Some of the more serious types of breast cancer, such as triple-negative breast cancer, account for 10.0%–20.8% of all the pathological types of breast cancer, with special biological behavior and clinicopathological characteristics, poor prognosis, and low survival rate compared with other types [14,15]. At present, although advanced treatment methods and early detection have improved survival rates, the causes of breast cancer remain unclear [12,14]. M5C modification has been shown to play an important role in cancer. Previous studies have found that m5C mRNA methylation is highly enriched in cancer-related pathways and NSUN6 expression is higher in healthy tissues than in tumors. NSUN6 is an m5C methyltransferase targeting mRNA, and its regulated methylation mainly occurs at 3′ UTRs near the downstream translation termination site of the stop codon, mainly affecting RNA and protein binding factors that regulate mRNA processing and translation in the testis [11]. High expression of NSUN6 in ovarian and liver tissues is associated with higher survival in patients with high fidelity of RNA and protein binding factors [12]. Using NSUN6 down-regulation as an marker in at least some cancers may be a new tumor biomarker [3].

## Functional role of NSUN6 in MCF7 cells

However, the research on breast cancer and m5C modification lacks substantial progress, and the relevant molecular regulatory mechanism remains to be discovered. As a regulatory factor related to m5C modification, NSUN6 has also been found to be abnormally expressed in breast cancer [5]. Some studies have found that NSUN6 is closely associated with the breast cancer stage. This suggests that the high expression of NSUN6 in tissues may also promote the high fidelity of RNA and protein binding factors involved in mRNA processing and translation, affecting the survival of breast cancer patients [12,14,16]. All these findings reflect that the regulatory mechanism of NSUN6 may have a strong relationship with breast cancer. In fact, this point was also proved by our experimental conclusion that up-regulation of NSUN6 promoted cell proliferation and migration, while down-regulation affected normal cell proliferation.

## Potential downstream targets and future directions

In our experiment, we constructed MCF7 cell lines with abnormal NSUN6 expression to observe the influence on growth and migration and obtained the changing trend of related modified genes by bioinformatic means. Our study found that changes in NSUN6 expression level in MCF7 cell lines up-regulated or down-regulated some genes, which ultimately affected normal cell growth and migration. Through RNA-seq, we analyzed all the related genes with change and finally confirmed about 10 genes with the same change trend as NSUN6. Among which EFEMP1 gene, previous studies have found that it can regulate the migration and differentiation of glial cells and the ability of glial cells to support neurite growth [19]. And TNS3 gene is located in the cytoplasm, participates in local adhesion, and acts in dephosphorylation and intracellular signal transduction acts upstream or within cell migration, and has a positive regulatory effect on cell proliferation [20]. FGFR2 gene as cancer also already had a lot of research [21]. These genes may also be found in subsequent breast cancer studies. These studies may be the future looking for specific NSUN6 modification of specific molecular pathways, and also help in diagnosis and treatment of patients, eventually improve the prognosis of patients.

## Observations in triple-negative cell lines

Besides, we also selected the other types of breast cancer cell lines for experiment but found that the growth of different breast cancer cells and phenotypic changes show a big difference. For example, some triple-negative breast cancer cells, such as M231, BT549, and M468, showed very inconsistent intra-group changes after NSUN6 knockdown, and the cell morphology and growth also showed irreversible changes. These experimental results were very unreliable, which may need to be paid attention to in subsequent relevant experiments. Why they show these results is also a very noteworthy problem.

## Conclusion

In summary, this study establishes NSUN6 as a critical driver of breast cancer progression through its m5C methyltransferase activity. We demonstrate that NSUN6 functionally promotes proliferation and migration in MCF7 cells and identify a downstream gene network—including EFEMP1, TNS3, and FGFR2—that likely mediates its oncogenic effects. These findings not only elucidate a novel epitranscriptomic mechanism in breast cancer but also nominate NSUN6 as a promising druggable epigenetic target. Our work provides a mechanistic foundation for further exploring NSUN6-specific pathways and supports the future development of epitranscriptome-directed therapeutic strategies for breast cancer.

## Ethics and Consent Statement

The MCF7 cell line was obtained from ATCC. All experimental procedures were performed in accordance with institutional biosafety guidelines

## Author Contributions

Song designed the study, performed all experiments, analyzed the data, and wrote the manuscript.

## Acknowledgments

This work was assisted by Southern University of Science and Technology, which was responsible for providing the experimental site, equipment and reagents.

## Funding

This study received no external funding.

## Disclosure

The author reports no conflicts of interest in this work.